\documentclass{article}
\usepackage{graphicx} \graphicspath{ {seipati/} }
\usepackage{amsmath}
\usepackage{color}
\usepackage{fancyhdr}
\usepackage{verbatim}
\usepackage{amsfonts,longtable}
\usepackage{epsfig}
\usepackage{amsfonts}
\usepackage{color}

\def\R{\hbox{{\rm I}\kern-0.2em{\rm R}\kern0.2em}}

\def\d{{\rm d}}

\def\ep{\epsilon}

\def\bn{\begin{equation}}
\def\en{\end{equation}}
\def\bny{\begin{eqnarray}}
\def\eny{\end{eqnarray}}
\def\be{\begin{eqnarray*}}
\def\ee{\end{eqnarray*}}
\def\bc{\begin{center}}
\def\ec{\end{center}}
 
\def\p{\partial} \def\q{{\bar q}}
\def\({\left(}
\def\){\right  )}
\def\[{\left[}
\def\]{\right]}
\def\bc{\begin{center}}
\def\ec{\end{center}}

\newtheorem{dfn}{Definition}[section]
\newtheorem{thm}{Theorem}[section]
\newtheorem{rem}{Remark}[section]
\newtheorem{pro}{Proposition}[section]

\newtheorem{cor}{Corollary}[section]
\newtheorem{lem}{Lemma}[section]
\newtheorem{exm}{Example}[section]

\def\bn{\begin{equation}}
\def\en{\end{equation}}
\def\bny{\begin{eqnarray}}
\def\eny{\end{eqnarray}}
\def\be{\begin{eqnarray*}}
\def\ee{\end{eqnarray*}}
\def\bdn{\begin{dfn}}
\def\edn{\end{dfn}}
\def\btm{\begin{thm}}
\def\etm{\end{thm}}
\def\bpf{\begin{proof}}
\def\epf{\end{proof}}
\def\bpn{\begin{pro}}
\def\epn{\end{pro}}
\def\brk{\begin{rem}}
\def\erk{\end{rem}}
\def\bcy{\begin{cor}}
\def\ecy{\end{cor}}
\def\blm{\begin{lem}}\def\elm{\end{lem}}
\def\bex{\begin{exm}}
\def\eex{\end{exm}}

 \def\R{{\cal R}}
\def\I{{\cal I}}

\begin{document}

\title{\textbf{Invariance and conservation laws of some nonlinear Schr\"odinger equation with \textbf{PT}-symmetric potentials
and inhomogeneous nonlinearity}}
\author{B Alqurashi and A H Kara\footnote{\footnotesize Corresponding author. E-mail address:
Abdul.Kara@wits.ac.za} \\
\\
 School of Mathematics, University of the Witwatersrand,\\ Private Bag 3, Wits 2050,
Johannesburg, South Africa}
\date{}
\maketitle
\begin{quote}

{\textbf{Abstract:} {\small In this paper, we construct and analyse the symmetries and conservation laws (conserved densities) of
a model of a nonlinear Scr\"odinger equation with \textbf{PT}-symmetric potentials and inhomogeneity.  
}}


\end{quote}

\section{Introduction}

In \cite{yan}, Yan et al establish and discuss, in detail, a model of a nonlinear Scr\"odinger equation with \textbf{PT}-symmetric (parity-time) potentials and inhomogeneity. The literature describing \textbf{PT}  models is vast, many of which are referred to in \cite{yan} and other works are \cite{a1,b1,c1} but we would allude the reader to the article on conservation laws and exact solution of certain \textbf{PT}-symmetric models in \cite{mkb}. In particular,  Yan et al consider `how stable nonlinear modes can be excited
in systems where the linear \textbf{PT} symmetry is broken. The
idea is based on the possibility of "switching on" nonlinearity
simultaneously with gain and dissipation. Such a possibility
can be implemented, in particular, when the nonlinearity and
gain-and-loss strength are characterized by a single parameter
$\ep$ and disappear when this parameter becomes
zero, i.e., $\ep=0$. If at $\ep=0$ the system is Hamiltonian (\emph{variational}), it allows for
stable propagation of the linear modes, and the only stability
issue which has to be verified is the stability of the solution
branch  $\ep>0$, bifurcating from $\ep=0$'.

The model developed and discussed in \cite{yan} (with $\mu=\ep$) is
\begin{equation}
iq_t+\frac12 u_{xx}-U_\ep(x)q-G_\mu(x)|q|^2q  = 0,
\label{1}\end{equation}
where
\begin{equation}
U_\ep(x)=\Sigma^{2}_{j=0}\ep^{2j}V_j(x)+i\ep W(x), \qquad G_\mu=\mu^2G(x).
\label{1.0}\end{equation}

We will show, in fact, that for $\ep>0$, the model is not variational and, therefore,
one cannot appeal to Noether's theorem to determine conservation laws. In fact, there are no conservation laws for this case.
Yet, the system does display non trivial symmetry properties. This would lead us to a novel concept (and consequent procedure) of `approximate/perturbed
conservation laws'. That is, if a system is variational for a parameter $\ep=0$ and not variational for $\ep>0$ (the perturbed system) but shares the symmetry properties that lead to conservation laws in the former case (via Noether's theorem, for example), then one may construct, for $\ep$ close to zero, approximate conservation laws for the perturbed system that are exact conservation laws in the limiting case $\ep\rightarrow 0$ for the unperturbed case. The notions of approximate symmetries, approximate variational symmetries and associated conservation laws have been discussed in the past, for example, \cite{bgi,ukm,k}. The approach of `partial Lagrangians' \cite{partial}  may also be a route to determining some sort of conservation laws.
Here, however, the concepts and methodology are quite different and rely on the result that the Euler operator annihilates total divergences. For our purposes, it would be almost zero, i.e., up to order $\ep$.

\section{Preliminaries}

We present some preliminaries that will be used in the analyses that follow.

Consider an $r$th-order system of partial differential equations (pdes)
of $n$ independent variables $\underbar{s}=(s_1,s_2,\ldots,s_n)$ and $m$
dependent variables $u=(u_1,u_2,\ldots,u_m) $ viz.,
\begin{equation}
	E(\underbar{s},u,u_{(1)},\ldots,u_{(r)})= 0,~~~ ~~~ ~~~u =1,\ldots,\tilde{m},
\end{equation}
where a locally analytic function $f(\underbar{s}, u, u_1,\dots,u_k)$ of a finite number of dependent variables $u, u_1,\dots,u_k$ denote the collections of all first , second ,$\dots$, $k$th-order partial derivatives and $s$ is a multivariable, that is
\begin{equation}
u_i^\alpha = D_i(u^\alpha), ~~~ ~~~ ~~~ u_{ij}^\alpha = D_jD_i(u^\alpha),\dots
\end{equation}
respectively, with the total differentiation operator with respect to $s^i$ given by,
\begin{equation}
D_i = \frac{\partial}{\partial s^i} + u_i^\alpha \frac{\partial}{\partial u^\alpha} + u_{ij}^\alpha \frac{\partial}{\partial u_j^\alpha}+\dots ~~~ ~~~ ~~~ i = 1,\dots ,m.
\end{equation}

In order to determine conserved densities and fluxes, we resort to
the invariance and multiplier approach based on the well known
result that the Euler-Lagrange operator annihilates a total
divergence. Firstly, if $(T^{s1}, T^{s1},\ldots)$ is a conserved
vector corresponding to a conservation law, then
\begin{equation} D_{s1} T^{s1}+D_{s1} T^{s1}+\ldots=0\label{n1}\end{equation}
along the solutions of the differential equation $E(\underbar{s},u,u_{(1)},\ldots,u_{(r)})= 0$.\\
\\
Moreover, if there exists a nontrivial differential function $Q$,
called a `multiplier', such that
\begin{equation}\label{3.5a}
	 Q (\underbar{s}, u, u_{(1)} \dots) E(\underbar{s},u,u_{(1)},\ldots,u_{(r)}) =D_{s1} T^{s1}+D_{s1} T^{s1}+\ldots ,
\end{equation}
for some (conserved) vector $(T^{s1}, T^{s1},\ldots)$, then
\begin{equation}\label{3.5b}
	{{\delta}\over {\delta u}}[ Q (\underbar{s}, u, u_{(1)} \dots) E(\underbar{s},u,u_{(1)},\ldots,u_{(r)})] = 0,
\end{equation}
where ${{\delta}\over {\delta u}}$ is the Euler operator. Hence, one may determine the multipliers, using (\ref{3.5b}) and then construct the corresponding conserved vectors; several approaches for this exists of
which the better known one is the `homotopy' approach.

If the system of differential equations is derived from a variational principle, then the conserved vector components are obtainable from Noether's Theorem which requires, firstly, the
construction of variational symmetries (vector fields) $X=\xi^{s^i}\frac{\partial}{\partial s^i} + \eta^{u^\alpha} \frac{\partial}{\partial u^\alpha}$
 that leave the action integral invariant.  It is well known that the vector fields that leave the system of differential equations invariant (generators of Lie point symmetries) contain the algebra of variational symmetries, if the latter exists \cite{o,i,s}.

Conservation laws may be expressed as conserved forms \cite{ian}. For example, if $\underbar{s}=(t,x)$, the conserved form would be
$$\omega=T^t\d x-T^x\d t$$ (where $(T^t,T^x)$ is the conserved vector such that $D_tT^t+D_xT^x=0$ on the solutions of the pde $E(x,t,u,u_{(1)},\ldots,u_{(r)})= 0$ ). Here, $T^t\d x$ leads to the `conserved density' if $t$ and $x$ are time and space, respectively.

\section{Conservation laws}

In what follows below, we construct conservation laws for cases that are modelled and listed in \cite{yan}. The significance of these cases are discussed therein.

\subsection{Case 1}

For most of the cases in \cite{yan}, it turns out that we may write $U_\ep(x)=a(x)+i\ep b(x)$. Then, if $q=u+iv$, separation of the real and imaginary parts in (\ref{1}) leads to the system

\begin{equation}\begin{array}{ll}
	&u_{{t}}+\frac12\,v_{{xx}}-\varepsilon b \left( x \right) u-a \left( x
 \right) v+2\,{\mu}^{2}\sigma\,{{\rm e}^{-\alpha\,{x}^{2}}} \left( {u}
^{2}+{v}^{2} \right) v=0, \\
&-v_{{t}}+\frac12\,u_{{xx}}-a \left( x \right) u+\varepsilon\,b \left( x
 \right) v+2\,{\mu}^{2}\sigma\,{{\rm e}^{-\alpha\,{x}^{2}}} \left( {u}
^{2}+{v}^{2} \right) u=0
\end{array}
\label{2}\end{equation}


The pde (\ref{1}) and, therefore, the system (\ref{2}) is not derivable from a variational principle and the conservation laws, therefore, cannot be obtained via Noether's theorem as is the usually the case for Schr\"odinger type equations. It can be shown further that the respective systems have no conservation laws. However, the decoupled equivalent system (\ref{2}) does
admit two symmetry generators $X_1=\p_t$ and $X_2=u\p_v-v\p_u$ which are usually associated with energy and charge conservation, respectively. The respective characteristics $Q_1=(v_t,u_t)$ and $Q_2=(u,-v)$ construed as multipliers do not satisfy (\ref{3.5b}), i.e.,
\begin{equation}\begin{array}{ll}\label{3.5bb}
&{{\delta}\over {\delta (u,v)}}[ Q_i^1( u_{{t}}+\frac12\,v_{{xx}}-\varepsilon b \left( x \right) u-a \left( x
 \right) v+2\,{\mu}^{2}\sigma\,{{\rm e}^{-\alpha\,{x}^{2}}} \left( {u}
^{2}+{v}^{2} \right) v       )\\
&+Q_i^2(-v_{{t}}+\frac12\,u_{{xx}}-a \left( x \right) u+\varepsilon\,b \left( x
 \right) v+2\,{\mu}^{2}\sigma\,{{\rm e}^{-\alpha\,{x}^{2}}} \left( {u}
^{2}+{v}^{2} \right) u  )]\\
& \neq \textbf{0}
\end{array}\end{equation}

If we regard $\epsilon$ as a `small' parameter, we may suppose (\ref{1}) as a perturbation of some system with $\epsilon=0$ in (\ref{1}). A number of situations may be pursued, as a result. For example, the notions of approximate symmetries and approximate conservation laws (and their possible associations) as expounded in \cite{bgi} and \cite{ukm} may be studied. However, proceeding with the multiplier approach on constructing conservation laws, we revisit (\ref{3.5bb}).
Below, we show that in fact the Euler operator equals $\ep\textbf{w}$ which goes to zero as $\ep$ goes to zero. Thus, we can construct conservation laws that are `approximate' upto an order of $\ep$. We enumerate,  below, some of the cases (as in \cite{yan}) of (\ref{2}).

(a). $a(x)=\frac12 x^2$, $b(x)=x$:

1. Energy

The Euler operator is
\begin{equation}\begin{array}{ll}
	&{{\delta}\over {\delta (u,v)}}[ v_t(u_{{t}}+\frac12\,v_{{xx}}-\varepsilon b \left( x \right) u-a \left( x
 \right) v+2\,{\mu}^{2}\sigma\,{{\rm e}^{-\alpha\,{x}^{2}}} \left( {u}
^{2}+{v}^{2} \right) v   )\\
&+u_t(-v_{{t}}+\frac12\,u_{{xx}}-a \left( x \right) u+\varepsilon\,b \left( x
 \right) v+2\,{\mu}^{2}\sigma\,{{\rm e}^{-\alpha\,{x}^{2}}} \left( {u}
^{2}+{v}^{2} \right) u  )]\\
&=(-2\ep x v_t  ,2\ep xu_t  )\rightarrow 0 \quad \text{as}   \quad \ep\rightarrow 0
\end{array}\end{equation}
and the corresponding `approximate' conserved form is
\bn\begin{array}{ll}
\omega^1&=[
1/5\,t\varepsilon\,xv_{{t}}u_{{}}-1/5\,t\varepsilon\,xu_{{t}}v_{{}}-{
\mu}^{2}\sigma\,{{\rm e}^{-\alpha\,{x}^{2}}}{v_{{}}}^{2}{u_{{}}}^{2}-1
/2\,{\mu}^{2}\sigma\,{{\rm e}^{-\alpha\,{x}^{2}}}{v_{{}}}^{4}-1/2\,{
\mu}^{2}\sigma\,{{\rm e}^{-\alpha\,{x}^{2}}}{u_{{}}}^{4}-1/4\,v_{{}}v_
{{xx}}\\
&+1/4\,{x}^{2}{v_{{}}}^{2}-1/4\,u_{{}}u_{{xx}}+1/4\,{x}^{2}{u_{
{}}}^{2}]\d x\\
&+[ -1/5\,{x}^{2}\varepsilon\,v_{{t}}u_{{}}+1/5\,{x}^{2}\varepsilon\,u_{{2
}}v_{{}}+1/4\,v_{{x}}v_{{t}}+1/4\,u_{{x}}u_{{t}}-1/4\,v_{{}}v_{{xt}}-
1/4\,u_{{}}u_{{xt}} ]\d t
\end{array}\en
so that the the conserved density is
$$\Phi^t=-\frac1{5}\ep tx\I(q\q_t)-\frac12\mu^2 {{\rm e}^{-\alpha\,{x}^{2}}}|q|^4+\frac14|q|^2-\frac14\R(q\q_{xx}) .$$

2. Charge

\begin{equation}\begin{array}{ll}
	&{{\delta}\over {\delta (u,v)}}[ v_t(u_{{t}}+\frac12\,v_{{xx}}-\varepsilon b \left( x \right) u-a \left( x
 \right) v+2\,{\mu}^{2}\sigma\,{{\rm e}^{-\alpha\,{x}^{2}}} \left( {u}
^{2}+{v}^{2} \right) v   )\\
&+u_t(-v_{{t}}+\frac12\,u_{{xx}}-a \left( x \right) u+\varepsilon\,b \left( x
 \right) v+2\,{\mu}^{2}\sigma\,{{\rm e}^{-\alpha\,{x}^{2}}} \left( {u}
^{2}+{v}^{2} \right) u  )]\\
&=(-2\ep x u  ,-2\ep x v  )  \rightarrow 0 \quad \text{as}   \quad \ep\rightarrow 0
\end{array}\end{equation}

\bn\begin{array}{ll}
\omega^2&=[1/5\,t\varepsilon\,x{u_{{}}}^{2}+1/5\,t\varepsilon\,x{v_{{}}}^{2}-1/2
\,{u_{{}}}^{2}-1/2\,{v_{{}}}^{2}  ]\d x \\
&+[ -1/5\,{x}^{2}\varepsilon\,{u_{{}}}^{2}-1/5\,{x}^{2}\varepsilon\,{v_{{}
}}^{2}+1/2\,u_{{}}v_{{x}}-1/2\,v_{{}}u_{{x}}  ]\d t
\end{array}\en

Thus, the conserved density in complex functional form is $$\Phi^t=(\frac15 t \ep x-\frac12)|q|^2.$$

(b). $a(x)=\frac12 x^2$, $b(x)= - \left( \alpha+3 \right) x{{\rm e}^{-1/2\, \left( \alpha+1 \right) {x
}^{2}}}  $:

In this case, we present only the conserved density as the complete conserved form is cumbersome; this is in fact clear from even the density alone.

1. Energy

The Euler operator is
\begin{equation}\begin{array}{ll}
	&{{\delta}\over {\delta (u,v)}}[ v_t(u_{{t}}+\frac12\,v_{{xx}}-\varepsilon b \left( x \right) u-a \left( x
 \right) v+2\,{\mu}^{2}\sigma\,{{\rm e}^{-\alpha\,{x}^{2}}} \left( {u}
^{2}+{v}^{2} \right) v   )\\
&+u_t(-v_{{t}}+\frac12\,u_{{xx}}-a \left( x \right) u+\varepsilon\,b \left( x
 \right) v+2\,{\mu}^{2}\sigma\,{{\rm e}^{-\alpha\,{x}^{2}}} \left( {u}
^{2}+{v}^{2} \right) u  )]\\
&=(2\,v_{{t}}\varepsilon\,x{{\rm e}^{-1/2\, \left( \alpha+1 \right) {x}^{
2}}} \left( \alpha+3 \right)
   , -2\,u_{{t}}\varepsilon\,x{{\rm e}^{-1/2\, \left( \alpha+1 \right) {x}^
{2}}} \left( \alpha+3 \right)
  )\rightarrow 0 \quad \text{as}   \quad \ep\rightarrow 0
\end{array}\end{equation}
and the conserved density, $T^t$ is given by
\begin{equation}\begin{array}{ll}
T^t&=-{{1}\over{4{{x}^{4} \left( \alpha+1 \right) ^{5/2}}  }}
      \, [ 4\,{\mu}^{2}\sigma\,{{\rm e}^{-\alpha\,{x}^{2}}}
{v_{{}}}^{4}{x}^{4}\sqrt {\alpha+1}\alpha+4\,{\mu}^{2}\sigma\,{{\rm e}
^{-\alpha\,{x}^{2}}}{u_{{}}}^{4}{x}^{4}\sqrt {\alpha+1}\alpha+2\,{\mu}
^{2}\sigma\,{{\rm e}^{-\alpha\,{x}^{2}}}{u_{{}}}^{4}{x}^{4}\sqrt {
\alpha+1}{\alpha}^{2}  \\
&+18\,t\varepsilon\,\sqrt {\pi }\sqrt {2}{\rm erf}
 \left(1/2\,\sqrt {2}x\sqrt {\alpha+1}\right)v_{{t}}u_{{}}+4\,{\mu}^{2
}\sigma\,{{\rm e}^{-\alpha\,{x}^{2}}}{v_{{}}}^{2}{u_{{}}}^{2}{x}^{4}
\sqrt {\alpha+1}+12\,t\varepsilon\,u_{{t}}v_{{}}\sqrt {\alpha+1}{x}^{3
}{{\rm e}^{-1/2\, \left( \alpha+1 \right) {x}^{2}}}    \\
&-36\,t\varepsilon\,
v_{{t}}u_{{}}\sqrt {\alpha+1}x{{\rm e}^{-1/2\, \left( \alpha+1
 \right) {x}^{2}}}-12\,t\varepsilon\,v_{{t}}u_{{}}\sqrt {\alpha+1}{x}^
{3}{{\rm e}^{-1/2\, \left( \alpha+1 \right) {x}^{2}}}+36\,t\varepsilon
\,u_{{t}}v_{{}}\sqrt {\alpha+1}x{{\rm e}^{-1/2\, \left( \alpha+1
 \right) {x}^{2}}}    \\
 &-18\,t\varepsilon\,\sqrt {\pi }\sqrt {2}{\rm erf}
\left(1/2\,\sqrt {2}x\sqrt {\alpha+1}\right)u_{{t}}v_{{}}+2\,{\mu}^{2}
\sigma\,{{\rm e}^{-\alpha\,{x}^{2}}}{v_{{}}}^{4}{x}^{4}\sqrt {\alpha+1
}{\alpha}^{2}+2\,{\mu}^{2}\sigma\,{{\rm e}^{-\alpha\,{x}^{2}}}{v_{{}}}
^{4}{x}^{4}\sqrt {\alpha+1}   \\
&+2\,{\mu}^{2}\sigma\,{{\rm e}^{-\alpha\,{x}
^{2}}}{u_{{}}}^{4}{x}^{4}\sqrt {\alpha+1}+v_{{}}v_{{xx}}{x}^{4}\sqrt
{\alpha+1}{\alpha}^{2}+2\,v_{{}}v_{{xx}}{x}^{4}\sqrt {\alpha+1}\alpha
+u_{{}}u_{{xx}}{x}^{4}\sqrt {\alpha+1}{\alpha}^{2}\\
&+2\,u_{{}}u_{{xx}}
{x}^{4}\sqrt {\alpha+1}\alpha-12\,t\varepsilon\,v_{{t}}u_{{}}\alpha\,
\sqrt {\alpha+1}x{{\rm e}^{-1/2\, \left( \alpha+1 \right) {x}^{2}}}-4
\,t\varepsilon\,v_{{t}}u_{{}}{\alpha}^{2}\sqrt {\alpha+1}{x}^{3}{
{\rm e}^{-1/2\, \left( \alpha+1 \right) {x}^{2}}}\\
&+4\,t\varepsilon\,u_{
{2}}v_{{}}{\alpha}^{2}\sqrt {\alpha+1}{x}^{3}{{\rm e}^{-1/2\, \left(
\alpha+1 \right) {x}^{2}}}+16\,t\varepsilon\,u_{{t}}v_{{}}\alpha\,
\sqrt {\alpha+1}{x}^{3}{{\rm e}^{-1/2\, \left( \alpha+1 \right) {x}^{2
}}}+4\,{\mu}^{2}\sigma\,{{\rm e}^{-\alpha\,{x}^{2}}}{v_{{}}}^{2}{u_{{}
}}^{2}{x}^{4}\sqrt {\alpha+1}{\alpha}^{2}\\
&+12\,t\varepsilon\,u_{{t}}v_{
{}}\alpha\,\sqrt {\alpha+1}x{{\rm e}^{-1/2\, \left( \alpha+1 \right) {
x}^{2}}}+6\,t\varepsilon\,\sqrt {\pi }\sqrt {2}{\rm erf} \left(1/2\,
\sqrt {2}x\sqrt {\alpha+1}\right)\alpha\,v_{{t}}u_{{}}-16\,t
\varepsilon\,v_{{t}}u_{{}}\alpha\,\sqrt {\alpha+1}{x}^{3}{{\rm e}^{-1/
2\, \left( \alpha+1 \right) {x}^{2}}}\\
&+8\,{\mu}^{2}\sigma\,{{\rm e}^{-
\alpha\,{x}^{2}}}{v_{{}}}^{2}{u_{{}}}^{2}{x}^{4}\sqrt {\alpha+1}\alpha
-6\,t\varepsilon\,\sqrt {\pi }\sqrt {2}{\rm erf} \left(1/2\,\sqrt {2}x
\sqrt {\alpha+1}\right)\alpha\,u_{{t}}v_{{}}\\
&-{x}^{6}{v_{{}}}^{2}\sqrt
{\alpha+1}-{x}^{6}{u_{{}}}^{2}\sqrt {\alpha+1}-{x}^{6}{v_{{}}}^{2}
\sqrt {\alpha+1}{\alpha}^{2}-2\,{x}^{6}{v_{{}}}^{2}\sqrt {\alpha+1}
\alpha-{x}^{6}{u_{{}}}^{2}\sqrt {\alpha+1}{\alpha}^{2}\\
&-2\,{x}^{6}{u_{{
}}}^{2}\sqrt {\alpha+1}\alpha+v_{{}}v_{{xx}}{x}^{4}\sqrt {\alpha+1}+u
_{{}}u_{{xx}}{x}^{4}\sqrt {\alpha+1} ]
	\end{array}\end{equation}

2. Charge

Similarly, the conserved density, $T^t$ is
\begin{equation}\begin{array}{ll}
T^t&={\frac {1}{2{x}^{4} \left( \alpha+1 \right) ^{5/2}}}[2\,{{\rm e}^{-1/2\, \left( \alpha+1 \right) {x}^{2}}}\sqrt {\alpha+1}{
\alpha}^{2}\epsilon\,t{x}^{3}{u_{{}}}^{2}+2\,{{\rm e}^{-1/2\, \left(
\alpha+1 \right) {x}^{2}}}\sqrt {\alpha+1}{\alpha}^{2}\epsilon\,t{x}^{
3}{v_{{}}}^{2}\\
&+8\,{{\rm e}^{-1/2\, \left( \alpha+1 \right) {x}^{2}}}
\sqrt {\alpha+1}\alpha\,\epsilon\,t{x}^{3}{u_{{}}}^{2}+8\,{{\rm e}^{-1
/2\, \left( \alpha+1 \right) {x}^{2}}}\sqrt {\alpha+1}\alpha\,\epsilon
\,t{x}^{3}{v_{{}}}^{2}+6\,{{\rm e}^{-1/2\, \left( \alpha+1 \right) {x}
^{2}}}\sqrt {\alpha+1}\epsilon\,t{x}^{3}{u_{{}}}^{2}\\
&+6\,{{\rm e}^{-1/2
\, \left( \alpha+1 \right) {x}^{2}}}\sqrt {\alpha+1}\epsilon\,t{x}^{3}
{v_{{}}}^{2}-{\alpha}^{2}{x}^{4}\sqrt {\alpha+1}{u_{{}}}^{2}-{\alpha}^
{2}{x}^{4}\sqrt {\alpha+1}{v_{{}}}^{2}+6\,{{\rm e}^{-1/2\, \left(
\alpha+1 \right) {x}^{2}}}\sqrt {\alpha+1}\alpha\,\epsilon\,tx{u_{{}}}
^{2}\\
&+6\,{{\rm e}^{-1/2\, \left( \alpha+1 \right) {x}^{2}}}\sqrt {
\alpha+1}\alpha\,\epsilon\,tx{v_{{}}}^{2}-3\,\epsilon\,t\sqrt {\pi }
\sqrt {2}{\rm erf} \left(1/2\,\sqrt {2}x\sqrt {\alpha+1}\right)\alpha
\,{u_{{}}}^{2}\\
&-3\,\epsilon\,t\sqrt {\pi }\sqrt {2}{\rm erf} \left(1/2
\,\sqrt {2}x\sqrt {\alpha+1}\right)\alpha\,{v_{{}}}^{2}-2\,\alpha\,{x}
^{4}\sqrt {\alpha+1}{u_{{}}}^{2}-2\,\alpha\,{x}^{4}\sqrt {\alpha+1}{v_
{{}}}^{2}\\
&+18\,{{\rm e}^{-1/2\, \left( \alpha+1 \right) {x}^{2}}}\sqrt
{\alpha+1}\epsilon\,tx{u_{{}}}^{2}+18\,{{\rm e}^{-1/2\, \left( \alpha+
1 \right) {x}^{2}}}\sqrt {\alpha+1}\epsilon\,tx{v_{{}}}^{2}-9\,
\epsilon\,t\sqrt {\pi }\sqrt {2}{\rm erf} \left(1/2\,\sqrt {2}x\sqrt {
\alpha+1}\right){u_{{}}}^{2}\\
&-9\,\epsilon\,t\sqrt {\pi }\sqrt {2}
{\rm erf} \left(1/2\,\sqrt {2}x\sqrt {\alpha+1}\right){v_{{}}}^{2}-{u_
{{}}}^{2}{x}^{4}\sqrt {\alpha+1}-{v_{{}}}^{2}{x}^{4}\sqrt {\alpha+1}
   ]
\end{array}\end{equation}

(c). $a(x)=\frac12 x^2$, $b(x)= -2\, \left( 2\, \left( \alpha+3 \right) {x}^{2}-\alpha-15 \right) x{
{\rm e}^{-1/2\, \left( \alpha+1 \right) {x}^{2}}}
$:

We only show here that the Euler operator is approximately zero so that one could obtain the approximate conservation law as above.

1. Energy

\begin{equation}\begin{array}{ll}
	&{{\delta}\over {\delta (u,v)}}[ u(u_{{t}}+\frac12\,v_{{xx}}-\varepsilon b \left( x \right) u-a \left( x
 \right) v+2\,{\mu}^{2}\sigma\,{{\rm e}^{-\alpha\,{x}^{2}}} \left( {u}
^{2}+{v}^{2} \right) v   )\\
&-v(-v_{{t}}+\frac12\,u_{{xx}}-a \left( x \right) u+\varepsilon\,b \left( x
 \right) v+2\,{\mu}^{2}\sigma\,{{\rm e}^{-\alpha\,{x}^{2}}} \left( {u}
^{2}+{v}^{2} \right) u  )]\\
&=(4\,\epsilon\,v_t \left( 2\,\alpha\,{x}^{2}+6\,{x}^{2}-\alpha-15 \right) x
{{\rm e}^{-1/2\, \left( \alpha+1 \right) {x}^{2}}}
   , - 4\,\epsilon\,u_t \left( 2\,\alpha\,{x}^{2}+6\,{x}^{2}-\alpha-15 \right) x
{{\rm e}^{-1/2\, \left( \alpha+1 \right) {x}^{2}}}   ) \\
&   \rightarrow 0 \quad \text{as}   \quad \ep\rightarrow 0
\end{array}\end{equation}

2. Charge

\begin{equation}\begin{array}{ll}
	&{{\delta}\over {\delta (u,v)}}[ u(u_{{t}}+\frac12\,v_{{xx}}-\varepsilon b \left( x \right) u-a \left( x
 \right) v+2\,{\mu}^{2}\sigma\,{{\rm e}^{-\alpha\,{x}^{2}}} \left( {u}
^{2}+{v}^{2} \right) v   )\\
&-v(-v_{{t}}+\frac12\,u_{{xx}}-a \left( x \right) u+\varepsilon\,b \left( x
 \right) v+2\,{\mu}^{2}\sigma\,{{\rm e}^{-\alpha\,{x}^{2}}} \left( {u}
^{2}+{v}^{2} \right) u  )]\\
&=(4\,\epsilon\, \left( 2\,\alpha\,{x}^{2}+6\,{x}^{2}-\alpha-15 \right) x
{{\rm e}^{-1/2\, \left( \alpha+1 \right) {x}^{2}}}u,
4\,\epsilon\, \left( 2\,\alpha\,{x}^{2}+6\,{x}^{2}-\alpha-15 \right) x
{{\rm e}^{-1/2\, \left( \alpha+1 \right) {x}^{2}}}v)\\
&\rightarrow 0 \quad \text{as}   \quad \ep\rightarrow 0
\end{array}\end{equation}

\subsection{Case 2}

In the case of a double-well potential with  \textbf{PT}-symmetry phases of the linear problem, we choose
$a(x,g)=1/2\,{x}^{2}-1/2\,{g}^{2}{{\rm e}^{-{x}^{2}}}-2\,\sigma\,{g}^{4}{
{\rm e}^{- \left( \alpha+1 \right) {x}^{2}}}$ and $b(x)=-3/2\,x{{\rm e}^{-{x}^{2}}}$ so that the Euler operator on the the following cases again yield
vectors that go to zero.

1. Energy

$${{\delta}\over {\delta (u,v)}}[\ldots]=(-3\,v_{{t}}\epsilon\,x{{\rm e}^{-{x}^{2}}},-3\,u_{{t}}\epsilon\,x{{\rm e}^{-{x}^{2}}})\rightarrow 0 \quad \text{as}   \quad \ep\rightarrow 0$$
\begin{equation}\begin{array}{ll}
T^t&={{1}\over{16x^4}}[8\,{{\rm e}^{-\alpha\,{x}^{2}}}{\mu}^{2}\sigma\,{u_{{}}}^{4}{x}^{4}+16
\,{{\rm e}^{-\alpha\,{x}^{2}}}{\mu}^{2}\sigma\,{u_{{}}}^{2}{v_{{}}}^{2
}{x}^{4}+8\,{{\rm e}^{-\alpha\,{x}^{2}}}{\mu}^{2}\sigma\,{v_{{}}}^{4}{
x}^{4}+16\,{u_{{}}}^{2}\sigma\,{g}^{4}{{\rm e}^{- \left( \alpha+1
 \right) {x}^{2}}}{x}^{4}\\
 &+16\,{v_{{}}}^{2}\sigma\,{g}^{4}{{\rm e}^{-
 \left( \alpha+1 \right) {x}^{2}}}{x}^{4}+4\,{u_{{}}}^{2}{g}^{2}{
{\rm e}^{-{x}^{2}}}{x}^{4}+4\,{v_{{}}}^{2}{g}^{2}{{\rm e}^{-{x}^{2}}}{
x}^{4}-12\,{{\rm e}^{-{x}^{2}}}\epsilon\,t{x}^{3}u_{{}}v_{{t}}+12\,{
{\rm e}^{-{x}^{2}}}\epsilon\,t{x}^{3}u_{{t}}v_{{}}\\
&-4\,{u_{{}}}^{2}{x}^
{6}-4\,{v_{{}}}^{2}{x}^{6}+9\,\epsilon\,t\sqrt {\pi }{\rm erf} \left(x
\right)v_{{t}}u_{{}}-9\,\epsilon\,t\sqrt {\pi }{\rm erf} \left(x
\right)u_{{t}}v_{{}}-18\,{{\rm e}^{-{x}^{2}}}\epsilon\,txu_{{}}v_{{t}}
+18\,{{\rm e}^{-{x}^{2}}}\epsilon\,txu_{{t}}v_{{}}\\
&+4\,u_{{}}u_{{xx}}{
x}^{4}+4\,v_{{}}v_{{xx}}{x}^{4}
]  , \\
T^x&= {{1}\over{16x^3}} [-12\,{{\rm e}^{-{x}^{2}}}\epsilon\,{x}^{3}u_{{}}v_{{t}}+12\,{{\rm e}^{
-{x}^{2}}}\epsilon\,{x}^{3}u_{{t}}v_{{}}+9\,\epsilon\,\sqrt {\pi }
{\rm erf} \left(x\right)v_{{t}}u_{{}}-9\,\epsilon\,\sqrt {\pi }
{\rm erf} \left(x\right)u_{{t}}v_{{}}\\
&-18\,v_{{t}}\epsilon\,x{{\rm e}^{
-{x}^{2}}}u_{{}}+18\,u_{{t}}\epsilon\,x{{\rm e}^{-{x}^{2}}}v_{{}}-4\,u
_{{xt}}u_{{}}{x}^{3}+4\,u_{{t}}u_{{x}}{x}^{3}-4\,v_{{xt}}v_{{}}{x}^{
3}+4\,v_{{t}}v_{{x}}{x}^{3}]

	\end{array}\end{equation}

2. Charge

$${{\delta}\over {\delta (u,v)}}[\ldots]=(3\,\epsilon\,x{{\rm e}^{-{x}^{2}}}u   ,3\,\epsilon\,x{{\rm e}^{-{x}^{2}}}v)\rightarrow 0 \quad \text{as}   \quad \ep\rightarrow 0$$
\begin{equation}\begin{array}{ll}
T^t&={{1}\over{16x^4}}[12\,{{\rm e}^{-{x}^{2}}}\epsilon\,t{x}^{3}{u_{{}}}^{2}+12\,{{\rm e}^{-
{x}^{2}}}\epsilon\,t{x}^{3}{v_{{}}}^{2}+18\,{{\rm e}^{-{x}^{2}}}
\epsilon\,tx{u_{{}}}^{2}+18\,{{\rm e}^{-{x}^{2}}}\epsilon\,tx{v_{{}}}^
{2}-9\,\epsilon\,t\sqrt {\pi }{\rm erf} \left(x\right){u_{{}}}^{2}\\
&-9\,
\epsilon\,t\sqrt {\pi }{\rm erf} \left(x\right){v_{{}}}^{2}-8\,{u_{{}}
}^{2}{x}^{4}-8\,{v_{{}}}^{2}{x}^{4}
]  , \\
T^x&=- {{1}\over{16x^3}} [-12\,{{\rm e}^{-{x}^{2}}}\epsilon\,{x}^{3}{u_{{}}}^{2}-12\,{{\rm e}^{-
{x}^{2}}}\epsilon\,{x}^{3}{v_{{}}}^{2}+9\,\sqrt {\pi }{\rm erf} \left(
x\right)\epsilon\,{u_{{}}}^{2}+9\,\sqrt {\pi }{\rm erf} \left(x\right)
\epsilon\,{v_{{}}}^{2}\\
&-18\,\epsilon\,x{{\rm e}^{-{x}^{2}}}{u_{{}}}^{2}
-18\,\epsilon\,x{{\rm e}^{-{x}^{2}}}{v_{{}}}^{2}+8\,v_{{x}}u_{{}}{x}^{
3}-8\,u_{{x}}v_{{}}{x}^{3}]
	\end{array}\end{equation}

The conserved density in complex function form is $\Phi^t={{1}\over{16x^4}}[12\,{{\rm e}^{-{x}^{2}}}\epsilon\,t{x}^{3}+18\,{{\rm e}^{-{x}^{2}}}
\epsilon\,tx-9\,\epsilon\,t\sqrt {\pi }{\rm erf} \left(x\right)-8\,{x}^{4}]|q|^2$.



\section{Conclusions}

We have shown that even though the class of Scr\"odinger equation with \textbf{PT}-symmetric potentials and inhomogeneity do not display variational properties and
are not conserved, one could construct quantities that are approximately conserved up to specified order of the `perturbation'.

\end{document}